\documentclass{article}
\usepackage[utf8]{inputenc}
\usepackage[T1]{fontenc}
\usepackage{authblk}
\usepackage{amssymb}
\usepackage{amsmath}
\usepackage{graphicx}
\usepackage[table,xcdraw]{xcolor}

\newtheorem{theorem}{Theorem}[section]

\newtheorem{remark}[theorem]{Remark}

\title{Towards a quantitative reduction of the SIR epidemiological model}

\author[1]{Matteo Colangeli\thanks{matteo.colangeli1@univaq.it}}
\author[2]{Adrian Muntean\thanks{adrian.muntean@kau.se}}
\affil[1]{Department of Information Engineering, Computer Science and Mathematics, University of L'Aquila, Via Vetoio, 67100 L'Aquila, Italy.}
\affil[2]{Department of Mathematics and Computer Science, Centre for Societal Risk Research (CSR), Karlstad University, Sweden.}

\begin{document}

\maketitle
\begin{abstract}

Motivated by our intention to use SIR-type epidemiological models in the context of dynamic networks as provided by large-scale highly interacting inhomogeneous human crowds, we investigate in this framework possibilities to reduce the classical SIR model to a representative evolution model for a suitably chosen observable. For selected scenarios, we provide practical {\em a priori} error bounds between the approximate and the original observables. Finally, we illustrate numerically the behavior of the reduced models compared to the original ones.
\end{abstract}
\section{Introduction}
The quest of a reduced description from a microscopic dynamics characterized by a large number of degrees of freedom is one of the classical problems of statistical and many-body physics, where model reduction and coarse-graining techniques proved to be a central tool underpinning renormalization group methods \cite{Kad,Wil}.
Recently, model reduction techniques also found relevant applications in meteorology \cite{Maj} and in physical and chemical kinetics \cite{coarse}.
One example is represented by the derivation of the hydrodynamic laws, described in terms of a restricted set of fields (e.g. density, momentum and temperature), from a kinetic description based on an extended set of moments or on the Boltzmann equation \cite{colan07a,colan07b,colan08,colan09}. With this occasion, we shall discuss the application of one such method of reduced description, called Invariant Manifold (IM) method \cite{GorKar05}, to the SIR model, which stands as one of historical benchmarks in the field of epidemic modelling. The approach can be adapted to our epidemiological  models described by coupled systems of nonlinear differential equations.  
As discussed in the sequel, the rationale behind the IM method is based on the identification of a restricted set of fields whose evolution, when observed in the appropriate time scale, captures some distinctive features of the microscopic dynamics of the system.  

This work belongs to the recent attempts of the applied mathematics community to understand, from a more fundamental perspective, the spread of viruses, like Covid-19, which are drastically affecting the well-being of our society; see e.g.  \cite{Britton, Fred,Perthame,Michael,Michael1,Tien}, citing but a few. Our own motivation stems from the potential use of SIR-type epidemiological models in the context of dynamic networks as provided by large-scale highly interacting inhomogeneous human crowds. In this framework, we identify possibilities to reduce the classical SIR model to a representative evolution model for a suitably chosen observable. 

\section{Basic SIR model and its quantitative reduction}
\label{sec:secSIR}
We focus our attention on the structure of the celebrated SIR model. We refer the reader, for instance, to \cite{Choisy} for a nice description of the modelling ideas behind SIR as well as to \cite{Michael,Michael1} for a number of qualitative properties of the solutions to SIR, SIRD, SEIR, and closely related models. 

We consider three populations of individuals belonging to a larger population whose total number of individuals is fixed. We shall denote by $S, I$, and $R$ the fraction of \textit{susceptible, infected,} and \textit{removed} individuals, respectively, such that $S+I+R=1$.

The model equations entering the SIR model are:

\begin{eqnarray}
\frac{d S}{d t}&=& -b~ S~ I,  \label{eqS} \\
\frac{d I}{d t}&=& b~ S~ I -\gamma~ I, \label{eqI} \\
\frac{d R}{d t}&=& \gamma~ I,  \label{eqR} 
\end{eqnarray}
where the initial conditions are prescribed as $S(0)=S_0,I(0)=I_0,R(0)=R_0$ such that $S_0+I_0+R_0=1$. Furthermore, $b$ and $\gamma$ are here strictly positive parameters that refer to an averaged infection rate constant and an averaged recovery rate constant, respectively. As a natural consequence, we see that if  $S_0+I_0+R_0=1$ holds, then we have that also the mass conservation law $S(t)+I(t)+R(t)=1$ holds for any $t\in (0,T)$, where $T>0$ is arbitrarily fixed.

We refer to the system of ODEs \eqref{eqS}--\eqref{eqR} as the \textit{original dynamics}. 

In this framework, we will offer a couple of other reduced variants of this SIR model. In all cases, we rely on the existence and uniqueness of classical positive solutions to the used models. 

To obtain a reduced description from the original dynamics, we make the following ansatz: we assume that a suitable time scale exists, in which the time evolution of $I(t)$ and $R(t)$ is driven by the dynamics of the \textit{leading} observable $S(t)$. 
 
 Under this assumption, we identify the so-called \textit{driven} observables. For the SIR model we have essentially two options:  either $\hat{S}(t)$ and $\hat{R}(t)$ or  $\hat{I}(t)$ and $\hat{R}(t)$, i.e. they correspond to  
\begin{equation}
    \hat{S}(t)=\Phi[\hat{I}(t)]  \quad \text{{\rm and}} \quad \hat{R}(t)=\Xi[\hat{I}(t)]. \label{ansatz}
\end{equation}
or 
\begin{equation}
    \hat{I}(t)=\Psi[\hat{S}(t)]  \quad \text{{\rm and}} \quad \hat{R}(t)=\Omega[\hat{S}(t)]. \label{ansatz2}
\end{equation}

The discussion of the reduction method based on Eq. \eqref{ansatz} is discussed in Sec. \ref{sec:secI}, while the analysis of \eqref{ansatz2} is deferred to Sec. \ref{sec:secS}.

\section{Using the constitutive law \(\hat{S}(t)=\Phi[\hat{I}(t)]\)}
\label{sec:secI}

The time evolution of the observables $\hat{S}(t)$, $\hat{I}(t)$ and $\hat{R}(t)$ is dictated by the original dynamics, Eqs. (\ref{eqS})--\eqref{eqR}, complemented by the ansatz (\ref{ansatz}). In particular, the dynamics of $\hat{I}(t)$ reads

\begin{equation}
\frac{d \hat{I}}{d t}= b~ \Phi[\hat{I}]~ \hat{I} -\gamma~ \hat{I}. \label{eqIred0} 
\end{equation}

Equation \eqref{eqSred0} corresponds to the desired \textit{reduced description} of the original SIR model, in which an expression for the constitutive law $\Phi[\hat{I}]$ is yet to be found. The dynamics of the driven observables $\hat{S}(t)$ and $\hat{R}(t)$ is given by

\begin{eqnarray}
    \frac{d \hat{S}}{d t}&=& -b~\Phi[\hat{I}(t)]~ \hat{I}   \label{eqSred0}\\
\frac{d \hat{R}}{d t}&=& \gamma~ \hat{I}.  \label{eqRred0} 
\end{eqnarray}

The initial value problem for the system \eqref{eqIred0}--\eqref{eqRred0} is defined by fixing the values $\hat{S}(0)=\hat{S}_0$, $\hat{I}(0)=\hat{I}_0$ and $\hat{R}(0)=\hat{R}_0$.
Furthermore, we also set $\Phi[\hat{I}(0)]=\Phi_0$ and $\Xi[\hat{I}(0)]=\Xi_0$.

We look for the exact expression of the functional $\Phi[\hat{I}]$, or, at least,  a good  approximate version thereof. To this aim, using the ansatz \eqref{ansatz}, we may also write the time derivative of the observable $\hat{S}(t)$ by relying on the chain rule, namely we have
\begin{equation}
\frac{d \hat{S}}{d t}=\frac{d \hat{I}}{d t}~ \Phi'[\hat{I}],  \label{eqSred02}
\end{equation}
with $\Phi'[\hat{I}]:=d \Phi[\hat{I}]/d \hat{I}$, 
whereas the time derivative of $\hat{I}$ is given by (\ref{eqIred0}).

The IM reduction method stipulates the equality of the two expressions of the time derivative of $\hat{S}(t)$ given in Eqs. (\ref{eqSred0}) and  (\ref{eqSred02}).
This procedure thus leads to the \textit{invariance equation} \cite{GorKar05,GorKar}, which reads

\begin{equation}
-b~\Phi[\hat{I}(t)] = \Phi'[\hat{I}](b~ \Phi[\hat{I}]~  -\gamma~).   \label{closS0} 
\end{equation}


While Eq. \eqref{closS0} attains an exact, although not explicit, solution in terms of the Lambert $W$ function \cite{Lambert}, we wish to follow here another route, and look for approximate, possibly explicit, solutions to Eq. \eqref{closS0}. We can then integrate the latter by separation of variables, thus obtaining
\begin{equation}
\log\frac{\Phi[\hat{I}(t)]}{\Phi_0}=\frac{b}{\gamma}(\Phi[\hat{I}(t)]-\Phi_0])+(\hat{I}(t)-\hat{I}_0)). \label{solS0}
\end{equation}

We note in passing that, by proceeding in the same manner with the observable $\hat{R}$, we obtain the corresponding invariance equation
\begin{equation}
    \Xi'[\hat{I}(t)]\left(\frac{b}{\gamma}~ \Phi[\hat{I}(t)] -1\right)=1.   \label{closR0} 
\end{equation}

Using now \eqref{closS0}, we can rewrite \eqref{closR0} in the form
\begin{equation}
    \Xi'[\hat{I}]=-\frac{\gamma \Phi'[\hat{I}(t)]}{b~ \Phi[\hat{I}(t)]}, \label{check01}
\end{equation}

which yields the expression
\begin{equation}
 \log \frac{\Phi[\hat{I}(t)]}{\Phi_0}=-\frac{b}{\gamma}\left[\Xi[\hat{I}(t)]-\Xi_0 \right]\label{check02}
\end{equation}

Finally, letting 
\begin{equation}
\Phi_0=\hat{S}_0 \quad \text{and} \quad
\Xi_0=\hat{R}_0 \label{initcon}
\end{equation}
and using \eqref{solS0} and \eqref{check02},  we obtain the consistency relation
\begin{equation}
    \hat{S(t)}+\hat{I(t)}+\hat{R(t)}=\hat{S}_0+\hat{I}_0+\hat{R}_0. \label{check03}
\end{equation}

which shows that the conservation of the total number of individuals in the population is inherited by the reduced description.

Next, in order to  determine an explicit approximate  expression of the functional $\Phi[\hat{I}]$, we seek approximate solutions of Eq. (\ref{solS0}) obtained via an iteration method. As a possible choice of iteration method, we propose
\begin{equation}
\log \frac{\Phi^{(i+1)}[\hat{I}(t)]}{\Phi_0}=\frac{b}{\gamma}\left[\left(\hat{I}(t)-\hat{I}_0\right)+\left(\Phi^{(i)}[\hat{I}(t)]-\Phi_0\right)  \right]    \label{solS02}
\end{equation}
For instance, by setting the initial condition for the recurrence equation (\ref{solS02}) equal to $\Phi^{(0)}[\hat{I}(t)]=\hat{I}(t)$, we find, at the the iteration level $i=1$, the solution
\begin{equation}
  \Phi^{(1)}[\hat{I}(t)]= \Phi_0 \exp\left\{\frac{b}{ \gamma}(2\hat{I}(t)-\hat{I}_0-\Phi_0)\right\}. \label{sol1}
\end{equation}
A natural question thus arises: What do we learn from approximate solutions? It will turn out that the better we can approximate the exact solution of Eq. \eqref{solS0}, given in terms of the Lambert $W$ function, the better we can reduce the SIR  model; see Claim 3 later on.


\subsection{A direct short-time estimate}
\label{sec:est1}

We now aim at estimating  the nearness of the solution $S(t)$ to Eq. (\ref{eqS}) with initial datum $S(0)=S_0$, and the \textit{constitutive law} $\Phi[\hat{I}(t)]$, defined in Eq. (\ref{ansatz}) with initial datum $\Phi[\hat{I}(0)]=\Phi_0$. It will turn out that our bound is meaningful only for a small time interval of observation and for a convenient parameter regime.

For an arbitrarily fixed value $\delta >0$ with $t^*\in (0,\delta )$, we can derive for any $t\in(0,\delta)$ the next upper bound:
\begin{eqnarray}
|S(t)-\Phi[I(t)]| &=& |S_0+\int_0^{t} \left(-bSI  \right)d\tau -\Phi[I(t)] |  \nonumber\\
&=& |S_0+\int_0^{t} \left(-bSI  \right)d\tau -\Phi\left[I_0+\int_0^{t} \left(bS I-\gamma I   \right)d\tau\right] |\nonumber\\
&\le& \mathcal{O}(\delta^2) + |S_0-\Phi[I_0]|
+|\left(-bS(t^*)I(t^*) \right)\delta| \nonumber\\
&+&  |\Phi'\left[I_0\right] \left(bS(t^*) I(t^*)-\gamma I(t^*)   \right)\delta  |. \nonumber\\
&\leq & |S_0-\Phi[I_0]|+ c^*\delta,\label{est18}
\end{eqnarray}
where $c^*>0$ is a constant depending on the parameters of the model, as well as on {\em a priori} uniform bounds on $S, I$ and on the smoothness of $\Phi$. 

The last term comes form the Taylor expansion of $\Phi[I(t)]$, while the term involving the point evaluation in $t^*$ is the result of the application of the mean-value theorem. Based on the rough estimate (\ref{est18}), we observe that the quantity 
$$e(t):=|S(t)-\Phi[I(t)]|$$
can be made small if $\delta>0$ is sufficiently small, $\Phi$ is at least twice differentiable and $S(t)$ and $I(t)$ are bounded positive continuous functions. Mind, however, that the smallness of $e(t)$ strongly depends on the choice of the parameters $b$ and $\gamma$. For instance, in the limit of large values of $\gamma$, 
most likely the quantity $e(t)$ will grow. On the other hand, if $\gamma$ takes moderate values, then we expect $e(t)\sim \mathcal{O}(\delta)$ for sufficiently small $t$ and $e(t)\sim \mathcal{O}(\delta T)$ for $t\in (0,T)$. 

\begin{remark} \begin{itemize}
    \item[(i)] The structure of the original ODE system indicates that if the functionals $\Phi[\cdot]$ and $\Xi[\cdot]$ are suitable exponentials (obtained by integrating the equations for $S$ and $R$), then $e(t)=0$ for any $t\in (0,T)$.
\item[(ii)] We expect that instead of estimating from above the quantity $e(t)$, it is more practical to bound the quantity  $|S(t)-\Phi[\hat I(t)]|$ for  $t\in(0,\tau)$, with $\tau$ fixed.
\end{itemize}
\end{remark}

\subsection{An indirect large-time estimate}
\label{sec:est2}

The dynamics of $I(t)$ is governed by 
\begin{equation}
    \frac{d I}{d t}= b~ S~ I -\gamma~ I,  \label{I0}
\end{equation}
where $I(0)=I_0$. 
The starting point of this discussion is the fact that besides
\begin{equation}
    \frac{d \hat I}{d t}= b~ \Phi[\hat I]~ \hat I -\gamma~ \hat I,  \label{I1}
\end{equation}
we may also consider
\begin{equation}
    \frac{d \tilde I}{d t}= b \Phi^{(n)}[\tilde I] \tilde I -\gamma \tilde I,  \label{I2}
\end{equation}
where $\Phi^{(n)}$ is the solution to our iterative method at the step $n\in\mathbb{N}$. We provide also the information on the initial data $I(0)$, $\hat I(0)$, and $\tilde I(0)$. 

{\bf Claim 1}: The iteration method works such that for any $r\in [0,||I||_\infty]$ it holds 
\begin{equation}
    |\Phi(r)-\Phi^{(n)}(r)|\leq \epsilon_n,
\end{equation}
with $\lim_{n\to\infty}\epsilon_n=0$. Here $||\cdot||_\infty$ denotes the standard uniform norm on $C[0,T]$. 

\paragraph{Proof of Claim 1.} Previous work done in the existing literature on the rigorous numerical approximation of the Lambert function gives trust in this Claim; see, for instance, \cite{Lambert} and references cited therein.

{\bf Claim 2}: Under the assumptions for which {\bf Claim 1} holds, there exist  constants $\hat c_1>0$ and $\hat c_2>0$ such that \begin{equation}|I(t)-\tilde I(t)|\leq e^{\hat c_1 T} \left(|I(0)-\tilde I(0)|+\hat c_2 T \epsilon_n\right)
\end{equation}
holds for any $t\in (0,T)$. Here $\hat c_1,\hat c_2$ are independent of $t,T$.

\paragraph{Proof of Claim 2.} We observe firstly that 
$$ |\Phi(I)- \Phi^{(n)}(\tilde I)|=|\Phi(I)- \Phi(\tilde I)+\Phi(\tilde I)-\Phi^{(n)}(\tilde I)|\leq|\Phi(I)- \Phi(\tilde I)|+|\Phi(\tilde I)-\Phi^{(n)}(\tilde I)|\leq $$
$$\leq ||\Phi'||_{\infty}|I-\tilde I|+ \epsilon_n.$$
Subtracting (\ref{I2}) from (\ref{I1}) gives:
$$\frac{d}{dt}(I-\tilde I)=\gamma (\tilde I-I)+b\left[\Phi(I)I-\Phi^{(n)}(\tilde I)\tilde I\right].$$
Noting that
$$\Phi(I)I-\Phi^{(n)}(\tilde I)\tilde I=I\left(\Phi(I)-\Phi^{(n)}(\tilde I)\right)+\Phi^{(n)}(\tilde I)(I-\tilde I)$$ 
leads to
$$\frac{d}{dt}|I-\tilde I|\leq (\gamma + |\Phi^{(n)}(\tilde I)|)|\tilde I-I|+||I||_\infty \left(||\Phi'||_{\infty}|I-\tilde I|+ \epsilon_n\right)\leq$$
$$\leq \epsilon_n||\Phi'||_{\infty}+ (\gamma +  |\Phi^{(n)}(\tilde I)|(1+||I||_\infty))|\tilde I-I|.$$
Now, using the Gr\"onwall's inequality (cf. e.g. Appendix B in \cite{Evans}) gives
$$|I(t)-\tilde I(t)|\leq e^{\int_0^t\left(\gamma+|\Phi^{(n)}(\tilde I(\tau))|(1+||I||_\infty)\right)d\tau}\left[|I(0)-\tilde I(0)|+||\Phi'||_\infty T\epsilon_n\right].$$
Choosing now $\hat c_1:=\gamma+||\Phi^{(n)}||_\infty|(1+||I||_\infty)$ and $\hat c_2:=||\Phi'||_\infty$ leads to the desired estimate proposed by Claim 2.

\subsection{Estimate on the error of the reduction method} 

In this section, we aim to bound from above the error produced by the reduction method proposed within this framework.

{\bf Claim 3}: Assume the hypothesis of Claim 2 to be true. Let $\tau>0$ be arbitrarily fixed. Then there exist strictly positive constants $\hat c_1, \hat c_2,$ and $\hat c_3$ such that  the following {\em a priori} estimate holds:
\begin{eqnarray}
\int_0^\tau|S(s)-\Phi(\hat I(s))|ds&\leq & \hat c_1 |I-\hat I|+ \hat c_2 \int_0^\tau |I(s)-\hat I(s)|ds\nonumber\\
&+&\hat c_3 |I(0)-\hat I(0)|,
\end{eqnarray}
where $I$ and $\hat I$ satisfy (\ref{eqI}),  and respectively (\ref{eqIred}). 

\paragraph{Proof of Claim 3.} 
We have
$$bS\hat I-b\Phi{(\hat I)}\hat I+Sb(I-\hat I)=\frac{d}{dt}(I-\hat I) +\gamma (I-\hat I).$$
By a direct manipulation of the structure of the equations (\ref{eqI}) and (\ref{eqIred}), we obtain:
$$S-\Phi{(\hat I)}=\frac{1}{b}\frac{1}{\hat I}\left[\frac{d}{dt}(I-\hat I) +\left(\gamma-S b \right)(I-\hat I)\right],$$
which by integration on $[0,\tau]$ leads to
\begin{eqnarray}\int_0^\tau|S(s)-\Phi{(\hat I(s))}|ds&\leq& \frac{1}{b}\int_0^\tau |\frac{1}{\hat I(s)}\left(\gamma-S(s)b \right)|ds\nonumber\\
&+&\frac{1}{b}|\int_0^\tau \frac{1}{\hat I(s)}\frac{d}{dt}(I(s)-\hat I(s))ds|.
\end{eqnarray}
As the first term on the right-hand side of the last inequality can be bounded above by $\left(\gamma+||S||_\infty b\right)\frac{1}{\beta}\int_0^\tau |I(s)-\hat I(s)|ds$ (with $0<\beta\leq I(t)$), and respectively, the last term by $\frac{2}{b}\left(|I(\tau)-\hat I(\tau)|+|I(0)-\hat I(0)|\right)$, the claim is now proven by choosing correspondingly the constants $\hat c_1$, $\hat c_3$,  and $\hat c_3$.
\begin{remark}
Combining the statements of Claim 3 and Claim 2, we note that it exists a constant $c>0$ such that
\begin{eqnarray}\label{23}
\int_0^\tau|S(s)-\Phi(\hat I(s))|ds&\leq & c\left(\epsilon_n+ |I(0)-\hat I(0)|\right).
\end{eqnarray}
This is obtained by adding and subtracting an $\tilde I$ in each term on the right-hand side  of the estimate provided by Claim 3 and employing conveniently the statement of Claim 2.
Note that the estimate (\ref{23}) is quite practical. It basically tells that if $I(0)=\hat I(0)$, then the quality of the reduction method depends mostly on the quality of the numerical approximation of the constitutive law $\Phi^n$.
\end{remark}
\section{Ideas for an improved reduced description}
\label{sec:reduced}
The analysis of the SIR model, developed in Sec. \ref{sec:secSIR}, has shed light on the conditions under which one may hope to quantitatively capture the features of the original dynamics by using a reduced description based on the IM method. We shall now turn our attention to another model, amenable to an analytical solution, which will also clarify the strengths and limitations of the IM method. 

\subsection{An instance of the Mori-Zwanzig method}
\label{sec:subMori}

The Mori-Zwanzig method, in its essence, performs a partition of the dynamical variables into two subsets corresponding to the ``relevant'' and the ``irrelevant'' variables. Suitable projection operators are then employed to project the original dynamics onto the subspace of the relevant variables \cite{zwanzig}.
The simplest case in which this method can be discussed is a linear system of coupled first order ODEs:
\begin{eqnarray}
\dot{x}&=&L_{11}x+L_{12}y \label{eqA}\\
\dot{y}&=&L_{21}x+L_{22}y \label{eqB}
\end{eqnarray}
with $x(0)=x_0$ and $y(0)=y_0$, and where $L_{ij}$, $i,j=1,2$, are real parameters.
Here $x(t)$ and $y(t)$ are regarded, respectively, as the ``relevant'' and the ``irrelevant'' variables. 

Using the set--up of Sec. \ref{sec:secSIR}, we may also regard $y(t)$ as the dynamical variable whose time evolution is driven by $x(t)$. 

In this case, the original dynamics, given by Eqs. \eqref{eqA}-\eqref{eqB}, is amenable to an analytical solution.
Namely, we can first solve (\ref{eqB}) for $y(t)$:
\begin{equation}
    y(t)=\exp\{L_{22}t\}y_0+\int_0^t \exp\{L_{22}(t-s)\}L_{21}x(s) ds, \label{yt}
\end{equation}
which represents the exact constitutive law linking $y(t)$ to $x(t)$. 
We can plug, next, (\ref{yt}) into (\ref{eqA}) to obtain a closed ODE for $x(t)$, which reads

\begin{equation}
    \dot{x}=L_{11} x+L_{12}\int_0^t \exp\{L_{22}(t-s)\}L_{21}x(s) ds + L_{12}\exp\{L_{22}t\}y_0 \ . \label{xt}
\end{equation}

Equation (\ref{xt}) represents the exact reduced description in terms of the relevant variable $x(t)$. The price we paid to get a reduced description, in this example, amounts to the presence of a memory term in the evolution equation for $x(t)$, which echoes the dynamics of the irrelevant variables.

We point out that, in the modelling of multiscale phenomena, the choice of the relevant dynamical variables is not always supported by guiding thermodynamic principles \cite{Grmela}. In fact, an improper choice of the relevant variables may not lead, eventually, to a successful reduced description \cite{Ott}. On the other hand, a meaningful selection of the relevant variables proved to be extremely important, in statistical mechanics, to establish general results such as the Fluctuation-Dissipation Relations and the Fluctuation Relations in nonequilibrium systems \cite{CRV,CR}.

\subsection{Using the Invariant Manifold method}
\label{sec:subIV}
 We shall now discuss the application of the IM method to the system \eqref{eqA}-\eqref{eqB}. 
To get started, we focus on a perturbative method known as the Chapman-Enskog expansion, which stems from the geometrical theory of singular perturbations \cite{Jones}.
Namely, we introduce a singular perturbation into the equation of $y(t)$ as follows:

\begin{equation}
    \dot{y}=L_{21}x+\frac{1}{\varepsilon}L_{22}y, \label{eqB2}
\end{equation}
where $\varepsilon>0$ is a small parameter.
Proceeding as in Sec. \ref{sec:secSIR}, we introduce the new variable $\hat{y}$ such that
\begin{equation}
    \hat{y}(t)=\Phi[\hat{x}(t)], \label{closure}
\end{equation}
where $\hat{x}(t)$ is defined via its ODE
\begin{equation}
    \dot{\hat{x}}=L_{11}~\hat{x}+L_{12}~\Phi[\hat{x}]. \label{xhat}
\end{equation}

The dynamics of $\hat{y}(t)$ can be described equivalently using the following two equations
\begin{eqnarray}
\frac{d \hat{y}}{d t}&=&L_{21}~x+\frac{1}{\varepsilon}L_{22}~\Phi[\hat{x}] \label{yhat1} \\
\frac{d \hat{y}}{d t}&=&\dfrac{d \hat{x}}{dt}~ \Phi'[\hat{x}].\label{yhat2}
\end{eqnarray}
We then impose the equality of two expressions of the time derivative of $\hat{y}(t)$, \eqref{yhat1} and \eqref{yhat2}, thus obtaining the invariance equation
\begin{equation}
  L_{21}~\hat{x}+\frac{1}{\varepsilon}L_{22}~\Phi[\hat{x}(t)]=\Phi'[\hat{x}]~ \left(L_{11}~\hat{x}+L_{12}~\Phi[\hat{x}]\right). \label{inv}
\end{equation} 

In the Chapman-Enskog method, the solution of Eq. \eqref{inv} is sought by expanding the variable $\hat{y}$ in a form of a series in powers of the number $\varepsilon$, i.e.
\begin{equation}
    \Phi[\hat{x}]=\Phi^{(0)}[\hat{x}]+\sum_{i=1}^{\infty}\varepsilon^i \Phi^{(i)}[\hat{x}]. \label{expan}
\end{equation}
Thus, the Eq. \eqref{inv} takes the form
\begin{eqnarray}
   &&L_{21}\hat{x}+\frac{1}{\varepsilon}L_{22}(\Phi^{(0)}+\varepsilon \Phi^{(1)}+ \varepsilon^2 \Phi^{(2)}+...)=\frac{d (\Phi^{(0)}+\varepsilon \Phi^{(1)}+ \varepsilon^2 \Phi^{(2)}+...)}{d \hat{x}} \times\nonumber\\
   &&\times [L_{11}\hat{x} + L_{12}(\Phi^{(0)}+\varepsilon \Phi^{(1)}+ \varepsilon^2 \Phi^{(2)}+...)], \label{inveq}
\end{eqnarray}
which must be solved order by order by equating terms on both sides of the equation. 
At the lowest orders of $\varepsilon$, one finds the following sequence of constitutive laws
\begin{eqnarray}
\text{order $\varepsilon^{-1}$:} \qquad  \Phi^{(0)}[\hat{x}]&=&0  \label{phi0} \\
\text{order $\varepsilon^{0}$:} \qquad  \Phi^{(1)}[\hat{x}]&=&-\frac{L_{21}}{L_{22}}\hat{x}   \label{phi1} \\
\text{order $\varepsilon$:} \qquad \Phi^{(2)}[\hat{x}]&=&-\frac{L_{21}L_{11}}{L_{22}^2}\hat{x}  \label{phi2} \\
\text{order $\varepsilon^2$:} \qquad \Phi^{(3)}[\hat{x}]&=&\frac{L_{21}}{L_{22}^3}(L_{12}L_{21}-L_{11}^2)\hat{x}.  \label{phi3}
\end{eqnarray}
An inspection of the structure of the approximated solutions in Eqs. (\ref{phi0})--(\ref{phi3}) gives us a hint on the structure of the solution of the invariance equation \eqref{inv} when no singular perturbation is introduced in Eq. \eqref{eqB2} (i.e. when setting $\varepsilon=1$ in Eq. \eqref{eqB2}) \cite{GorKar}.
Therefore, we seek for a constitutive law \eqref{closure} endowed with the linear structure
\begin{equation}
   \Phi[\hat{x}]=A~ \hat{x}, \label{exclos}
\end{equation}
  where $A(L_{11},L_{12},L_{21},L_{22})$ is an unknown function of the parameters $L_{ij}$, $i,j=1,2$, yet to be determined.
  Hence \eqref{inv} takes the form
  \begin{equation}
    L_{21}~\hat{x}+L_{22}~A~\hat{x}=A~(L_{11}~\hat{x}+L_{12}~A~\hat{x}). \label{inveq2}  
  \end{equation}
 Equation (\ref{inveq2}) becomes the quadratic equation
 \begin{equation}
     L_{12}~A^2+(L_{11}-L_{22})~A-L_{21}=0. \label{inveq3}
 \end{equation}
  whose roots are
  \begin{equation}
      A^{*}=\frac{-(L_{11}-L_{22})\pm \sqrt{(L_{11}-L_{22})^2+4L_{12}L_{21}}}{2L_{12}}. \label{solinv}
  \end{equation}
  Thus, the IM method leads to the following constitutive law 
  \begin{equation}
      \hat{y}=A^{*}\hat{x} \label{soly}
  \end{equation}
  which should be
  compared with the exact constitutive law \eqref{yt}.
  
\begin{figure}[t]
\centering
\includegraphics[width = 0.47\textwidth]{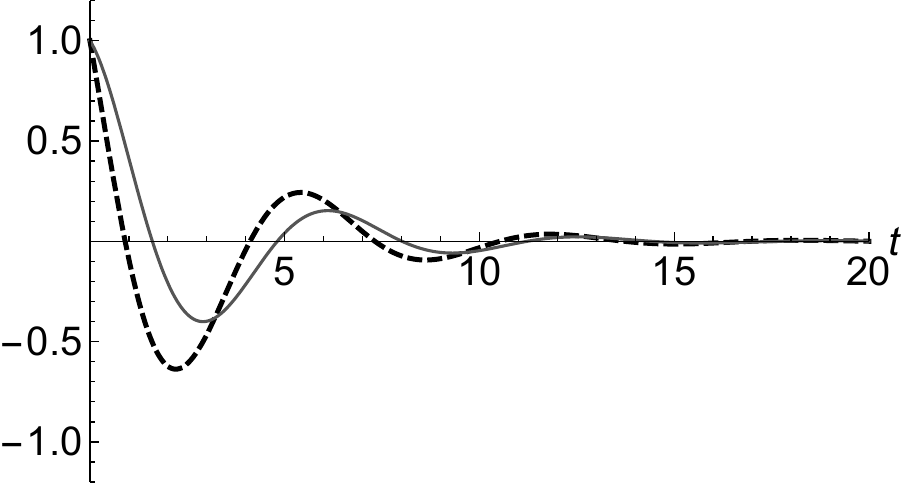}
\hspace{3mm}
\includegraphics[width = 0.47\textwidth]{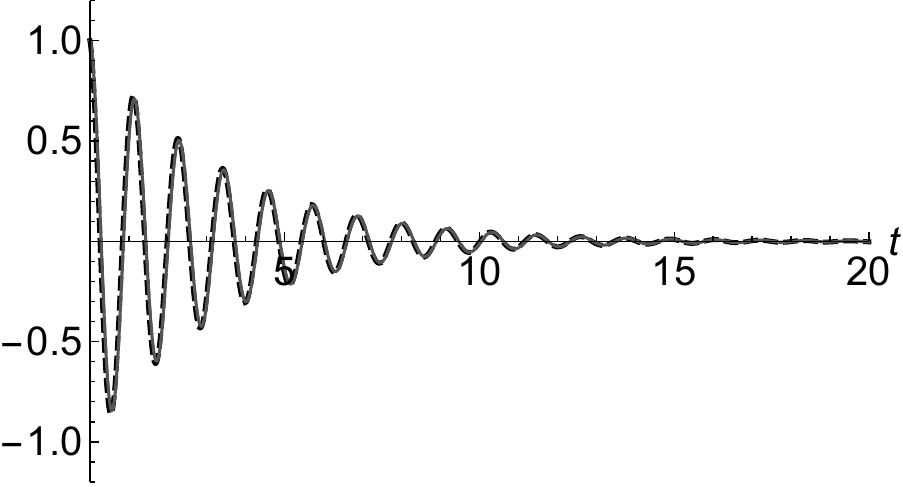}
\caption{\textit{Left panel:} behavior of $x(t)$ (black dashed line), obtained from Eq. \eqref{xt}, and $\Re(\hat{x}(t))$ (gray solid line), obtained from Eq. (\ref{redIV}), with $x_0=\hat{x}_0=1$, $L_{11}=-0.1$, $L_{12}=-1$ and $L_{22}=-0.5$ and with $L_{21}=1$ (left panel) and $L_{21}=30$ (right panel).}
\label{fig:fig3}
\end{figure}

  Using (\ref{soly}) we can now rewrite \eqref{xhat} as follows
  \begin{equation}
      \dot{\hat{x}}=\mathcal{A}^{*}\hat{x}, \label{redIV}
  \end{equation}
  where we have set $\mathcal{A}^{*}=L_{11}+L_{12}A^{*}$.

Figure \ref{fig:fig3} shows the behavior of $x(t)$, obtained by integration of Eq. \eqref{xt}, and of the real part $\Re(\hat{x}(t))$ of the solution of the reduced system (\ref{redIV}), with initial datum $x_0=\hat{x}_0=1$, for two different values of the coupling parameter $L_{21}$. In particular, Fig. \ref{fig:fig3} evidences that fixing a larger value of $L_{21}$, while keeping the other parameters fixed, leads to a better performance of the IM reduction method. The reason is that a larger value of $L_{21}$, in \eqref{eqB}, makes the time derivative of $y(t)$ more strongly affected by the behavior of $x(t)$. 
This, hence, fits nicely with the ansatz \eqref{closure}, which requires the dynamics of $\hat{y}(t)$ to be driven by $\hat{x}(t)$. 
  
  
  \subsection{Back to the SIR model}
  \label{sec:secS}

In this concluding section, we return to the SIR model of Sec. \ref{sec:secSIR} and discuss the application of the IM method by using the ansatz \eqref{ansatz2}. It will turn out that, in this case, the IM method may yield an exact reduced description. 


The dynamics of the field $\hat{S}(t)$ now reads

\begin{equation}
    \frac{d \hat{S}}{d t}= -b~ \hat{S}~ \Psi[\hat{S}(t)]  \label{eqSred}
\end{equation}
with $\hat{S}(0)=\hat{S}_0$. We also fix $\Psi[\hat{S}(0)]=\Psi_0$ and $\Omega[\hat{S}(0)]=\Omega_0$.

In the present case, Eq. \eqref{eqSred} corresponds to the reduced description of the original SIR model, Eqs. \eqref{eqS}--\eqref{eqR}. To find an explicit expression for the constitutive law $\Psi[\hat{S}]$, we write, first, the dynamics of the driven observable $\hat{I}(t)$ as

\begin{equation}
\frac{d \hat{I}}{d t}= b~ \hat{S}~ \Psi[\hat{S}(t)] -\gamma~ \Psi[\hat{S}(t)]. \label{eqIred} 
\end{equation}


Next, as in Sec. \ref{sec:secI}, we also write the time derivative of $\hat{I}(t)$ by using the chain rule, i.e.

\begin{equation}
\frac{d \hat{I}}{d t}=\frac{d \hat{S}}{d t}~ \Psi'[\hat{S}].  \label{eqIred2} 
\end{equation}

We thus obtain the invariance equation:


\begin{equation}
   \Psi'[\hat{S}(t)]=-1+\frac{\gamma}{b~\hat{S}(t)}. \label{closI2} 
\end{equation}

 We can then integrate Eq. (\ref{closI2}) by separation of variables, thus obtaining
\begin{equation}
\Psi[\hat{S}(t)]=\Psi_0-(\hat{S}(t)-\hat{S}_0)+\frac{\gamma}{b}\log\frac{\hat{S}(t)}{\hat{S}_0}.    \label{solI}
\end{equation}

\begin{figure}[ht]
\centering
\includegraphics[width = 0.6\textwidth]{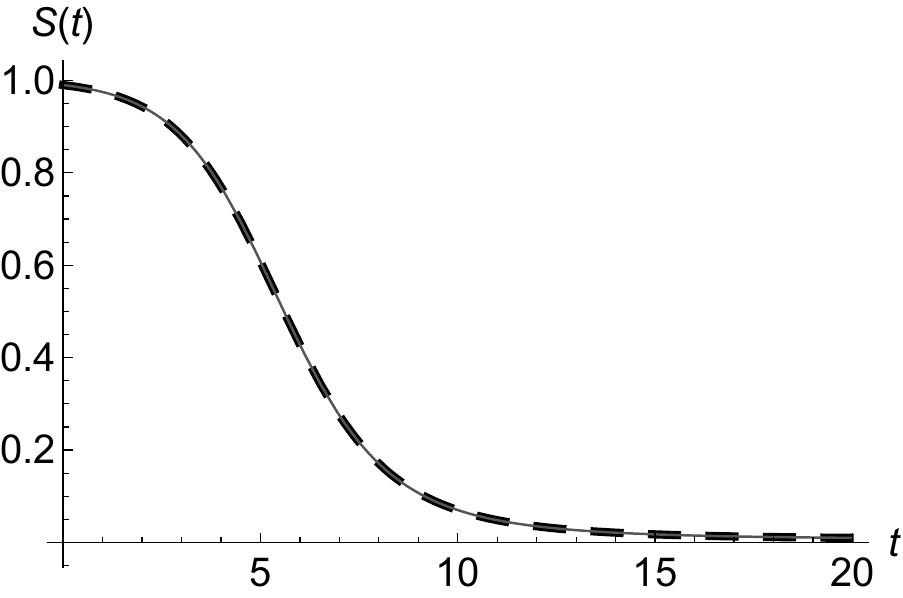}
\caption{Behavior of $S(t)$ for the original dynamics, Eq. \eqref{eqS}  (black dashed line), and for the reduced description (gray solid line), obtained from Eqs. (\ref{eqSred}) and (\ref{solI}). We fixed $S_0=\hat{S}_0=0.99$ and $\Psi_0=\hat{I}_0=0.01$, with $b=1$ and $\gamma=0.2$.}
\label{fig:fig1}
\end{figure} 

\begin{figure}[ht]
\centering
\includegraphics[width = 0.47\textwidth]{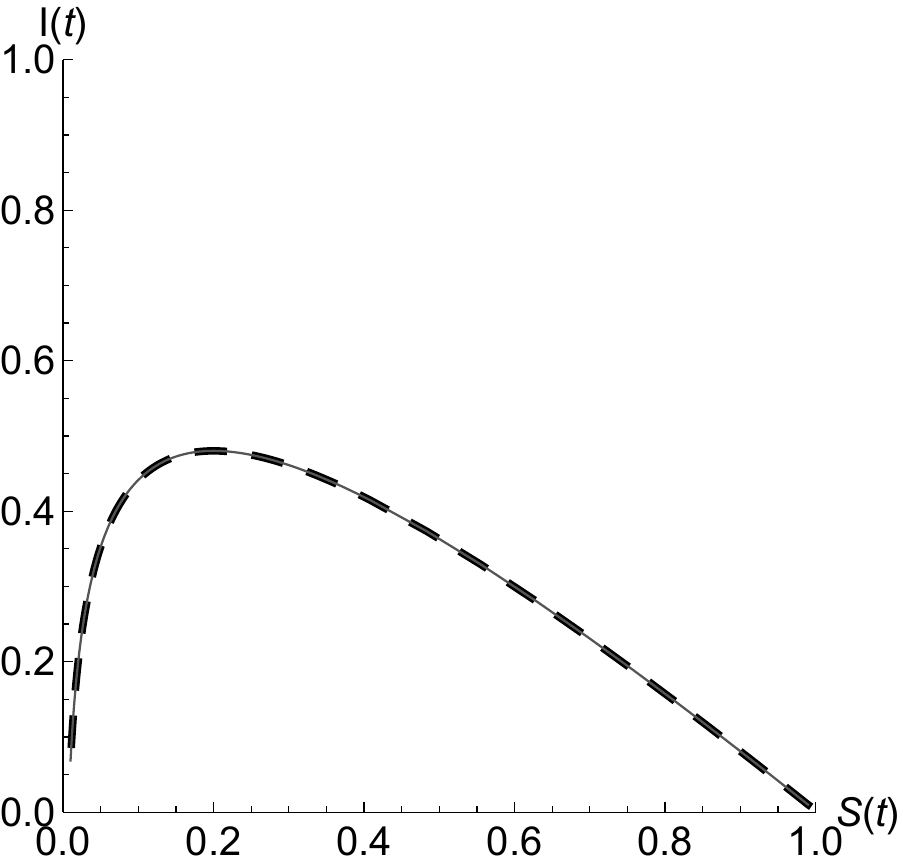}
\hspace{3mm}
\includegraphics[width = 0.47\textwidth]{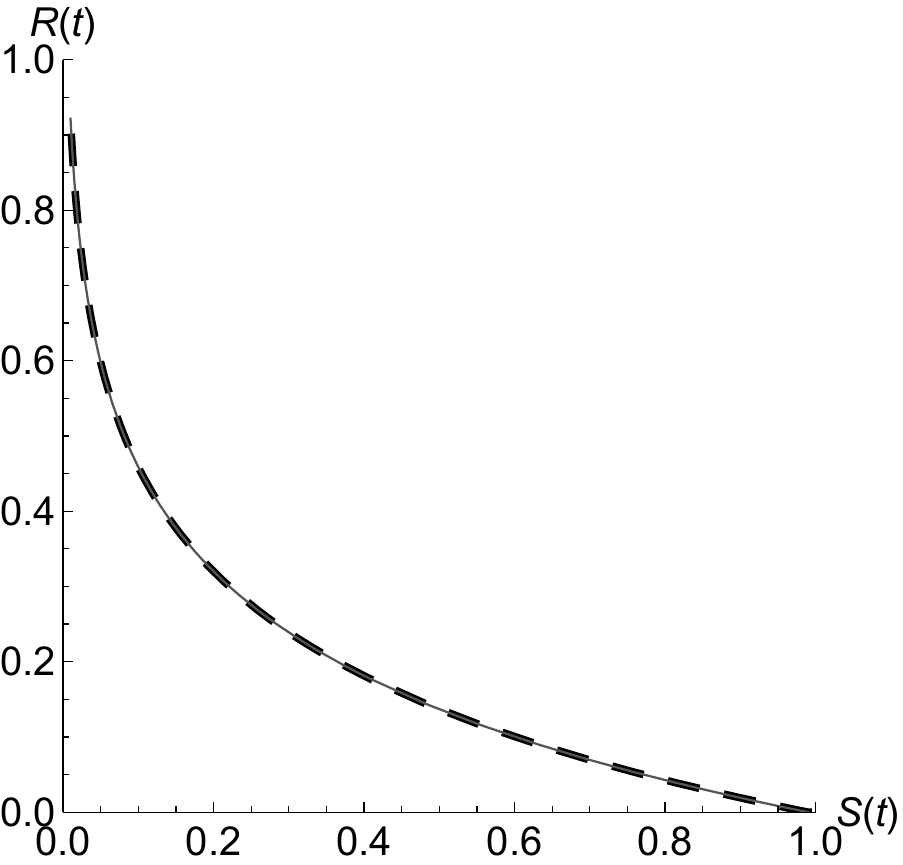}
\caption{\textit{Left panel:} parametric plots of $I(t)$ vs $S(t)$ for the original dynamics,  Eqs. (\ref{eqS})-\eqref{eqI} (black dashed line) and for the reduced description (gray solid line), obtained from Eq. \eqref{solI}. \textit{Right panel:} parametric plot of $R(t)$ vs $S(t)$ for the original dynamics (black dashed line) and for the reduced description (gray solid line), obtained from Eq. \eqref{solR}. We fixed $I_0=\hat{I}_0=\Psi_0=0.01$ and $R_0=\hat{R}_0=\Omega_0=0$.}
\label{fig:fig2}
\end{figure}


In a similar fashion, we find 
\begin{equation}
\Omega[\hat{S}(t)]=\Omega_0-\frac{\gamma}{b}\log\frac{\hat{S}(t)}{\hat{S}_0}.
  \label{solR}
\end{equation}

Finally, by setting

\begin{equation}
\Psi_0=\hat{I}_0 \quad \text{and} \quad
\Omega_0=\hat{R}_0 \label{initcon2}
\end{equation}

and by summing up \eqref{solI} and \eqref{solR}, we obtain 
\begin{equation}
   \hat{S}(t)+\hat{I}(t)+\hat{R}(t)=\hat{S}_0+\hat{I}_0+\hat{R}_0, \label{check}
\end{equation}
which yields, again, the conservation of the total number of individuals in the reduced dynamics.
In Fig. \ref{fig:fig1} the behavior of $S(t)$, obtained by integration of the original SIR model, Eq. (\ref{eqS})--\eqref{eqR}, is compared with the solution $\hat{S}(t)$ of Eq. \eqref{eqSred}, equipped with the constitutive law (\ref{solI}). Figure \ref{fig:fig1} shows that the behavior of $\hat{S}(t)$ recovers with striking accuracy that of $S(t)$.
Moreover, the two panels of Fig. \ref{fig:fig2} show the parametric plots of the $I(t)$ vs. $S(t)$ (left panel) and $R(t)$ vs. $S(t)$ (right panel) for the original SIR model and for the reduced description. 

We would like, however, to point out that the nice agreement between original and reduced dynamics outlined above might easily be lost when considering time dependent parameters $b=b(t)$ and $\gamma=\gamma(t)$, for $t>0$. A careful rewriting of the invariance equation 
is demanded to handle such case. We will discuss this scenario elsewhere. 

\section{Conclusion}

We have succeeded to identify constitutive laws to reduce either the presence of the population of susceptibles or of the infectives in the standard SIR model. The reduced descriptions, obtained using the IM method, agree via numerical simulations and practical {\em a priori} error bounds with what is expected from the original SIR dynamics. 

Our work opens the possibility to use the reduced SIR dynamics for reading-off data available, for instance, on demonstrated Covid-19 infections and deaths and, based on a parameter identification approach done at this level, produce a new forecast on the effects of the pandemic evolution.

From a long-term research perspective, the method discussed in these notes indicates new routes to be exploited to obtain reduced descriptions in yet uncharted, or only partially explored, territories, such as the  mathematical modelling of crowd dynamics \cite{CCM,thoa1,thoa2,thoa3} 
and uphill diffusions \cite{CC,CDP,CGGV}, in the framework of interacting particle systems.


\begin{thebibliography}{SK}

\bibitem{Perthame}
\textsc{L. Almeida, P.-A. Bliman, G. Nandin, B. Perthame, N. Vauchelet}, \emph{Final size and convergence rate for an epidemic in heterogeneous populations}. M3AS, 2021.

\bibitem{Britton}
\textsc{T. Britton, F. Ball, P. Trapman}, \emph{A mathematical model reveals the influence of population heterogeneity on herd immunity to SARS-CoV-2}. Science \textbf{369}, 6505: 846--849 (2020).


\bibitem{CC}
\textsc{E.N.M. Cirillo, M. Colangeli}, \emph{Stationary uphill currents in locally perturbed zero-range processes}, Phys. Rev. E \textbf{96}(5) 052137 (2017).

\bibitem{CCM}
\textsc{E.N.M. Cirillo, M. Colangeli, A. Muntean}, \emph{Effects of communication efficiency and exit capacity on fundamental diagrams for pedestrian motion in an obscure tunnel -- a particle system approach}, Multiscale Model. Sim. \textbf{14}(2), 906--922 (2016).

\bibitem{thoa1}
\textsc{E.N.M. Cirillo, M. Colangeli, A. Muntean, T.K.T. Thieu}, \emph{A lattice model for active - Passive pedestrian dynamics: A quest for drafting effects}, Math. Biosci. Eng., \textbf{17}(1), 460--477 (2020).

\bibitem{thoa2}
\textsc{E.N.M. Cirillo, M. Colangeli, A. Muntean, T.K.T. Thieu}, \emph{When diffusion faces drift: Consequences of exclusion processes for bi-directional pedestrian flows}, Physica D \textbf{413}, 132651 (2020).

\bibitem{Choisy}
\textsc{M. Choisy, J.-F.  Gu\'egan, P. Rohani}, \emph{Mathematical modeling of infectious diseases dynamics}. In {\em Encyclopedia of Infectious Diseases}, M. Tibayrenc (Ed.), 379--404, (2007).

\bibitem{CDP}
\textsc{M. Colangeli, A. De Masi, E. Presutti},
\emph{Microscopic models for uphill diffusion}, J. Phys. A Math. Theor. \textbf{50}, 435002 (2017).

\bibitem{CGGV}
\textsc{M. Colangeli, C. Giardin\`{a}, C. Giberti, C. Vernia}, \emph{Nonequilibrium two-dimensional Ising model with stationary uphill diffusion}, Phys. Rev. E \textbf{97}(3) 030103 (2018).

\bibitem{colan07a}
\textsc{M. Colangeli, I. V. Karlin, M. Kr\"{o}ger},
\emph{From hyperbolic regularization to exact hydrodynamics for linearized Grad's equations}, Phys. Rev. E \textbf{75}, 051204 (2007).

\bibitem{colan07b}
\textsc{M. Colangeli, I. V. Karlin, M. Kr\"{o}ger},
\emph{Hyperbolicity of exact hydrodynamics for three-dimensional linearized Grad's equations}, Phys. Rev. E \textbf{76}, 022201 (2007).

\bibitem{colan09}
\textsc{M. Colangeli, M. Kr\"{o}ger, H. C. \"{O}ttinger},
\emph{Boltzmann equation and hydrodynamic fluctuations}, Phys. Rev. E \textbf{80}, 051202 (2009).


\bibitem{CRV} 
\textsc{M. Colangeli, L. Rondoni, A. Vulpiani}, 
\emph{Fluctuation-dissipation relation for chaotic non-Hamiltonian systems}, J. Stat. Mech. L04002 (2012).

\bibitem{CR} 
\textsc{M. Colangeli, L. Rondoni}, 
\emph{Equilibrium, fluctuation relations and transport for irreversible deterministic dynamics}, Physica D \textbf{241}, 681--691 (2012).

\bibitem{thoa3}
\textsc{M. Colangeli, A. Muntean, O. Richardson, T.K.T. Thieu}, \emph{Modelling interactions between active and passive agents moving through heterogeneous environments}, Modeling and Simulation in Science, Engineering and Technology, 211--257, Springer Basel (2018).

\bibitem{Maj}
\textsc{D. T. Crommelin, A. J. Majda}, \emph{Strategies for model reduction: Comparing different optimal bases}, Journal of the Atmospheric Sciences \textbf{61}, 2206--2217 (2004).

\bibitem{Evans}
\textsc{L. C. Evans}, \emph{Partial Differential Equations}, 2nd Edition, Graduate Studies in Mathematics, vol. \textbf{19}, American Mathematical Society (2010). 

\bibitem{Lambert}
\textsc{M. Fasi, N.J. Higham and B. Iannazzo}, \emph{An algorithm for the matrix Lambert $W$ function}, SIAM J. Matrix Anal. Appl., {\bf 36}, No. 2, 669--685.

\bibitem{GorKar05}
\textsc{A. N. Gorban, I. V. Karlin}, \emph{Invariant Manifolds for Physical and Chemical Kinetics}, Lect. Notes Phys. \textbf{660}, Springer, Berlin (2005).

\bibitem{GorKar}
\textsc{A. N. Gorban, I. V. Karlin}, \emph{Hilbert's $6$th problem: Exact and approximate manifolds for kinetic equations}, Bulletin of the American Mathematical Society,
Volume \textbf{51} (2), 187--246 (2013).

\bibitem{coarse}
\textsc{A. N. Gorban, I. G. Kevrekidis, C. Theodoropoulos, N. K. Kazantzis, H. C. \"{O}ttinger}, \emph{Model Reduction and Coarse-Graining Approaches
for Multiscale Phenomena}, Springer, Berlin (2006).

\bibitem{Grmela}
\textsc{M. Grmela}, \emph{Role of thermodynamics in multiscale physics}, Comp. Math.  App. \textbf{65}, no. 10, 1457--1470 (2013).

\bibitem{Jones}
\textsc{C. K. R. T. Jones}, \emph{Geometric Singular Perturbation Theory}, Dynamical systems (Montecatini Terme, 1994), Lecture Notes in Math., vol. 1609, Springer, Berlin, 1995.

\bibitem{Kad}
\textsc{L.P. Kadanoff}, \emph{Scaling laws for Ising models near $T_c$}, Physics \textbf{2}, 263--272 (1966).


\bibitem{colan08}
\textsc{I. V. Karlin, M. Colangeli, M. Kr\"{o}ger},
\emph{Exact linear hydrodynamics from the Boltzmann equation}, Phys. Rev. Lett. \textbf{100}, 214503 (2008).

\bibitem{Tien}
 \textsc{W. R. KhudaBukhsh, S. K. Khalsa,  E. Kenah,  G. A. Rempala, J. H. Tien}, \emph{COVID-19 dynamics in an Ohio prison}, medRxiv 2021.01.14.21249782.
 
\bibitem{Ott} 
\textsc{H. C. \"{O}ttinger},
\emph{Beyond Equilibrium Thermodynamics}, John Wiley \& Sons (2005).

\bibitem{Fred}
\textsc{F.J. Vermolen}, \emph{A spatial Markov chain cellular automata model for the spread of viruses}.  preprint. arXiv:2004.05635 (2020)

\bibitem{zwanzig}
\textsc{R. Zwanzig}, \emph{Nonequilibrium Statistical Mechanics}, Oxford University Press (2001).

\bibitem{Wil}
\textsc{K.G. Wilson}, \emph{Renormalization group and critical phenomena I. Renormalization group and the Kadanoff scaling picture}, Phys. Rev. B \textbf{4}, 3174?-3183 (1971).

\bibitem{Michael}
\textsc{M. Wolff}, \emph{On build-up of epidemiologic models--Development of a $SEI^3RSD$ model for the spread of SARS-CoV-2}, Zeitschrift f\"ur angewandte Mathematik und Mechanik (ZAMM) \textbf{100} (11), 1--36 (2020).

\bibitem{Michael1}
\textsc{M. Wolff}, \emph{Mathematische Bemerkungen zu einem epidemiologischen Modell im Zusammenhang mit der Corona-Pandemie}, preprint, May 2020, DOI: 10.13140/RG.2.2.20564.35209.


\end{thebibliography}
\end{document}